\documentclass[reprint]{revtex4-1}

\usepackage{array}
\usepackage{graphicx}
\usepackage{verbatim}
\usepackage{color}
\usepackage{multirow}

\newcommand{\ket}[1]{\left\vert{#1}\right\rangle}

\begin{document}

\title{Composite Toffoli gate with two-round error detection}
\author{Cody Jones}
\email{ncodyjones@gmail.com}
\affiliation{Edward L. Ginzton Laboratory,
         Stanford University,
         Stanford, California 94305-4088, USA}

\begin{abstract}
We introduce a fault-tolerant construction to implement a composite quantum operation of four overlapping Toffoli gates.  The same construction can produce two independent Toffoli gates.  This result lowers resource overheads in designs for quantum computers by more than an order of magnitude.  The procedure uses Clifford operations and 64 copies of the non-Clifford gate $T = \exp[i \pi (I - \sigma^z) /8]$.  Quantum codes detect errors in the circuit.  When the dominant source of error is $T$-gate failure with probability $p$, then the composite Toffoli circuit has postselected failure rate of $3072p^4$ to lowest order.  
\end{abstract}

\maketitle

\section{Introduction}
Fault-tolerant quantum computing is the research effort to make quantum computers reliable despite the many ways that quantum hardware suffers from errors beyond an experimenter's control.  Techniques from physics, information theory, and computer science are employed to develop robust quantum processors.  The development of quantum error correction was a critical result for quantum computing in general, because it showed that arbitrarily complex computations could be executed on hardware with nonzero error rate~\cite{Calderbank1996,Steane1996,Preskill1998,Nielsen2000,Knill2005}. However, optimism was tempered by the realization that the \emph{resource overhead} (the redundancy in hardware that enables error correction) could be several orders of magnitude larger than a noise-free circuit for plausible error rates and interesting quantum algorithms~\cite{Knill2005,Isailovic2008,Jones2012,Fowler2012b}.  Current hardware designs can control fewer than ten quantum bits~\cite{Ladd2010,Lucero2012,Blatt2012,Politi2009,Maurer2012,Shulman2012}, so the million-qubit devices that implement fault-tolerant computation must be several technology generations away from the current state of the art.  To bridge the gap, research in fault-tolerant quantum computing focuses on developing methods to reduce the overhead and to perform reliable quantum computing on hardware that is simpler to design and fabricate.

This work addresses the most resource-intensive component in most, if not all, quantum computations.  An important result from quantum error correction is that, in any quantum code, there always exists one crucial operation that is not natively available~\cite{Zeng2007,Eastin2009}, and hence it is expensive to prepare.  A commonly selected operation is the Toffoli gate~\cite{Barenco1995,Jones2013,Eastin2012}, defined by $U_{\mathrm{Tof}}\ket{a,b,c} = \ket{a,b,c \oplus ab}$, where $(a,b,c)$ are binary variables and operation $\oplus$ is binary XOR.  This paper introduces a fault-tolerant construction for the Toffoli gate which can substantially lower the resource overhead in fault-tolerant quantum computing.  The two-round error detection in this paper is an improvement over the one-round error detection in Refs.~\cite{Jones2013,Eastin2012}.  By incorporating this construction into recent analyses of fault-tolerant quantum architectures~\cite{Jones2012,Fowler2012b}, we anticipate that the resource costs determined therein could be reduced by more than an order of magnitude.

The paper is organized as follows.  Section~\ref{preliminaries_section} outlines some notation and preliminary assumptions.  Section~\ref{overview_section} summarizes the error-detection methods implemented in the paper.  Section~\ref{error_detection_section} gives an explicit quantum-circuit procedure for producing a composite Toffoli gate.  Section~\ref{analysis_section} calculates the suppressed error probability that results from this construction.  Section~\ref{teleportation_section} shows how to make use of the composite Toffoli gate with teleportation.  Section~\ref{discussion_section} discusses the impact of these results on fault-tolerant quantum computing.

\section{Preliminaries}
\label{preliminaries_section}
An important distinction in this paper is made between quantum gates that are ``easy'' and ``hard.''  An operation is easy when it  has a direct and low-overhead implementation within a chosen error-correcting code.  Traditionally, these operations were labeled ``transversal,'' because they could be applied element-wise to a code block or in matching pairs element-wise between two code blocks, which ensured fault tolerance~\cite{Nielsen2000}.  However, modern codes like surface codes~\cite{Raussendorf2007,Fowler2009} do not actually use transversal gates.  Still, the distinction is important because some operations are hard, meaning they require substantially more overhead to perform.  Often, the hard operations invoke many easy operations to perform some distillation procedure~\cite{Knill2004,Bravyi2005,Meier2012,Bravyi2012,Jones2012c}.

In the important family of CSS codes~\cite{Calderbank1996,Steane1996}, as well as many more stabilizer codes~\cite{Gottesman1997}, the easy operations are Clifford gates.  The group of Clifford gates includes the Pauli operators $\sigma^x \equiv X$, \emph{etc.}.  The group is generated by the phase gate $S = \exp[i \pi (I - Z)/4]$, Hadamard $H = (1/\sqrt{2})(X+Z)$, and CNOT.  For convenience, we will also consider initialization and measurement in the $X$ and $Z$ bases to be easy operations, so we may say they are ``Clifford'' although they are not unitary.

By contrast, non-Clifford gates tend to be much more difficult.  References~\cite{Zeng2007,Eastin2009} show that there is always one operation required for universal quantum computing that is not transversal in a given code.  In the surface code~\cite{Raussendorf2007,Fowler2009}, only a subset of the Clifford group is natively available, while the rest must be ``injected'' into the code space. Injection is not a fault-tolerant process, so the injected states must be purified of errors, which is costly.  Distilling the non-Clifford operation $T = \exp[i \pi (I - Z) /8]$ requires about $50\times$ the circuit resources as a fault-tolerant CNOT~\cite{Fowler2012d}.  This disparity motivates our efforts to find a more efficient non-Clifford operation in the form of Toffoli gates.  Moreover, the high cost of fault-tolerant non-Clifford gates (compared to Clifford gates) is the justification for another assumption, that only errors in the non-Clifford $T$~gates are considered.

This paper derives quantum circuits in a way that is well-suited to surface code error correction.  The features of the surface code make some logical code operations more convenient than others.  In particular, only CNOT, Hadamard, and $X$- and $Z$-basis initialization and measurement are natively available~\cite{Fowler2009,Fowler2012}.  We later demonstrate an ancilla-aided $Y$-basis measurement.  Rotations by angles $\pi/2$ or $\pi/4$ about the $X$ and $Z$ axes on the Bloch sphere are possible but more costly, as they require magic-state distillation.  We assume that the non-Clifford $T$~gate (rotation by $\pi/4$ around the $Z$ axis) is available to produce logical Toffoli gates with error detection.

\section{Overview of main results}
\label{overview_section}
\begin{figure}
  \centering
  \includegraphics[width=5cm]{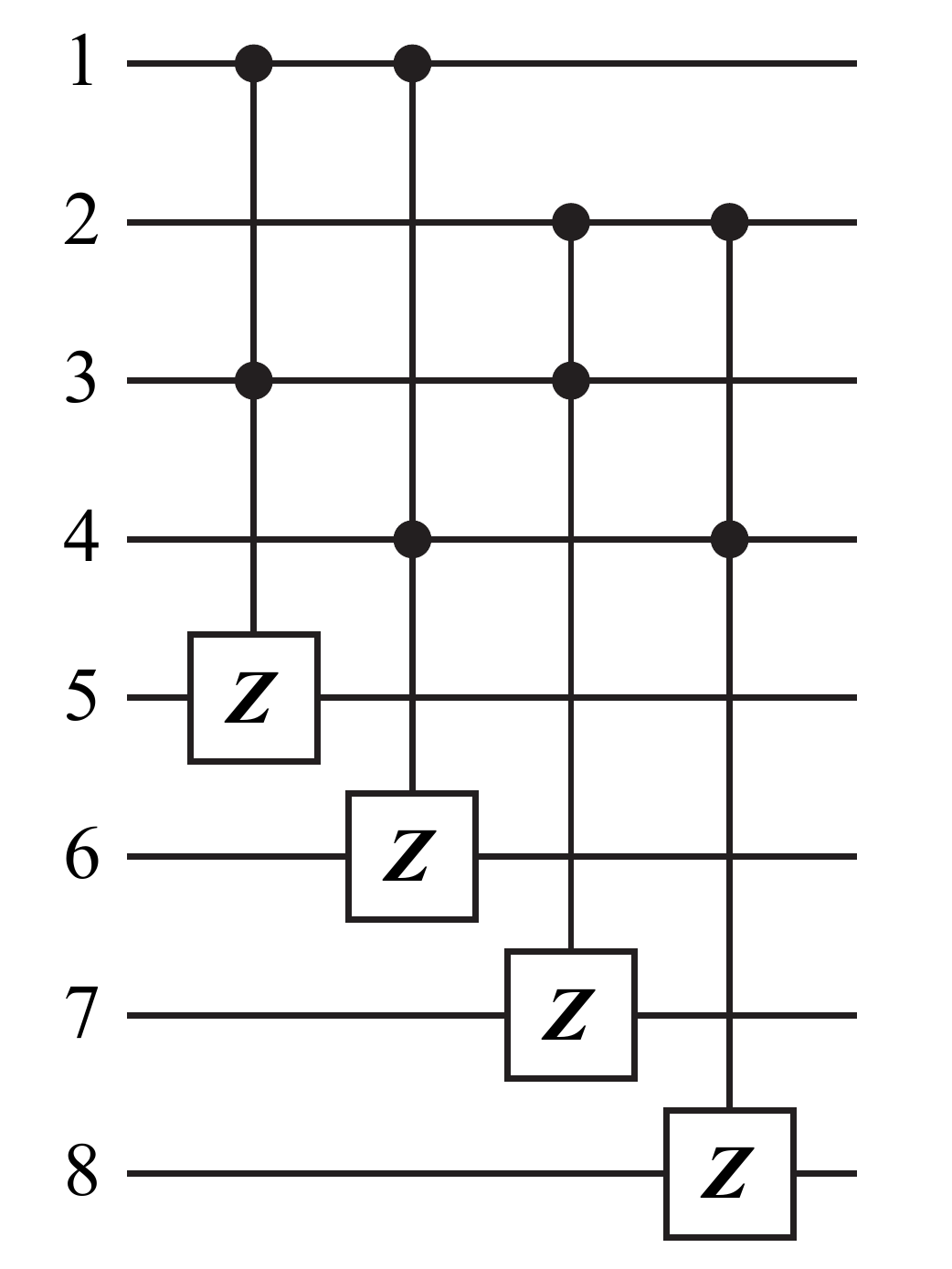}\\
  \caption{A composite CCZ gate acting on eight qubits, which are numbered for later reference.  Any qubit line could be converted to Toffoli target(s) using Hadamard gates, because CCZ is symmetric in its inputs.}
  \label{composite_CCZ_logical}
\end{figure}

We briefly summarize the main points of the composite-Toffoli construction to show what the analysis in later sections accomplishes. This high-level description is also useful for reference.  A form of the composite Toffoli gate is shown in Fig.~\ref{composite_CCZ_logical}, which depicts four controlled-controlled-$Z$ (CCZ) gates.  This circuit has flexibility to turn any particular qubit line into Toffoli-gate target(s) using Hadamard gates, which are local and Clifford.  Throughout most of the paper our approach is to create the composite-CCZ gate in Fig.~\ref{composite_CCZ_logical} and to assume that the appropriate Hadamard gates are inserted when this gate is used in an algorithm.  An important constraint to note is that these CCZ gates are inseparable, meaning they must all be implemented in the shown arrangement, without inserting any gates in the middle of the circuit.

The composite Toffoli is constructed with two rounds of error detection.  For now, we consider the only source of failure to be $T$~gates having $Z$ errors, each with independent probability $p$.  This simplifies the analysis and allows us to focus on the non-Clifford gates, which previous investigations found to be the most resource-costly component of fault-tolerant quantum computing~\cite{Jones2012,Fowler2012b,Fowler2012d,Fowler2013}.  Each round uses the $C_4$ error-detecting code~\cite{Knill2005}, which has distance two and which can detect a single error on any qubit.  The composite construction with two rounds of $C_4$ error detection has distance four with respect to $T$~gates.  As a result, the distance-four circuit will have postselected error of $O(p^4)$.

\begin{figure}
  \centering
  \includegraphics[width=8.3cm]{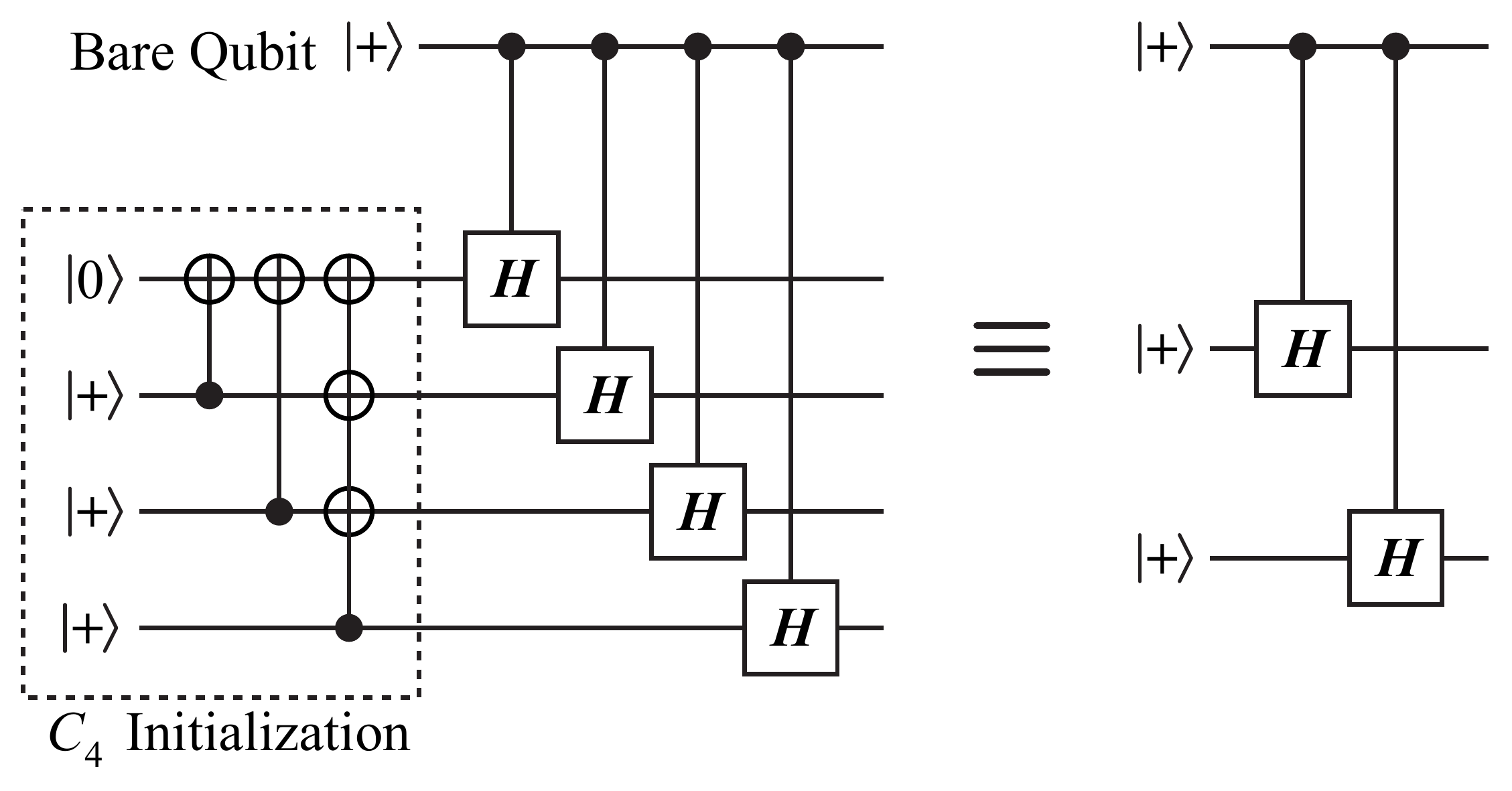}\\
  \caption{Circuit for constructing the three-qubit register encoding two coupled controlled-$H$ gates.  The right-hand side shows the equivalent logical circuit.}
  \label{C4_controlled_H_simple}
\end{figure}

In the first round of error detection, a $C_4$ code enables the construction of magic states for the controlled-$S$ gate.  The gate $S$ is a Clifford gate, but its controlled version is a non-Clifford gate from which Toffoli can be constructed.  The initial state of the error-detection circuit consists of a bare $\ket{+} = (1/\sqrt{2})(\ket{0} + \ket{1})$ qubit and a $C_4$ code block with two $\ket{+}$ encoded qubits.  The magic-state preparation will use four controlled-$H$ gates produced using eight $T$~gates, as we explain later.  Because transversal $H$ is a logical operation in $C_4$, the controlled-$H$ with the control on the bare qubit is also logical with respect to the code block~\cite{Meier2012}, as shown in Fig.~\ref{C4_controlled_H_simple}.  There are several important steps needed to make this process successful, and the procedure is detailed in Sec.~\ref{error_detection_section}.  Stabilizer measurements will detect a single error in the code block, and we later show that this will detect a single error in any of eight $T$~gates used in this procedure.  The output is a three-qubit magic state that can be used to produce two controlled-$S$ gates with a common control qubit (or common target, as controlled-$S$ is a symmetric operation).  The error probability for this three-qubit state is $28p^2$.

\begin{figure}
  \centering
  \includegraphics[width=8.3cm]{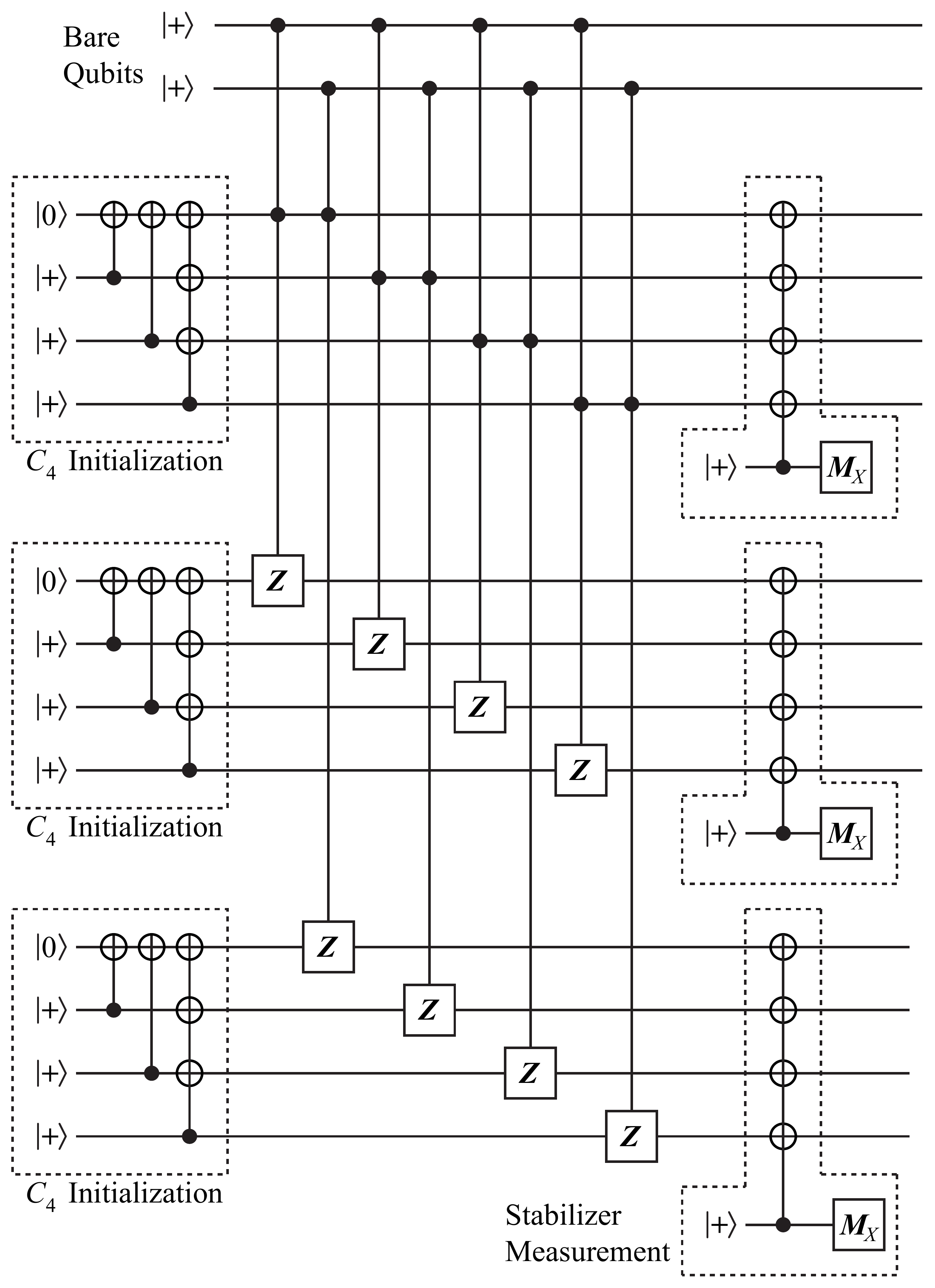}\\
  \caption{A high-level depiction of the second round of error detection in the composite-Toffoli circuit.  Each of the CCZ gates requires two controlled-$S$ gates produced using magic states from the first round shown in Fig.~\ref{C4_controlled_H_simple}.  The common control line from each coupled pair of controlled-$S$ gates is aligned with the top $C_4$ block.  The correspondence with Fig.~\ref{composite_CCZ_logical} is as follows: the bare qubits are inputs $(1,2)$; the pair of encoded qubits in each $C_4$ block are inputs $(3,4)$, $(5,6)$, and $(7,8)$, from top to bottom.}
  \label{composite_CCZ_overview}
\end{figure}

The second round of error detection also uses $C_4$ code blocks.  Transversal controlled-$Z$ is a logical operation between two $C_4$ codes.  Similar to above, we implement CCZ with a bare qubit controlling a transversal controlled-$Z$ operation between $C_4$ codes.  CCZ gates are constructed using controlled-$S$ gates, which are supplied by the magic states from above.  The controlled-$S$ gates act on the $C_4$ blocks, and a single error in any controlled-$S$ gate in each code block can be detected using the stabilizers of all $C_4$ code blocks.  To ensure independence of errors, any pair of controlled-$S$ gates with common control must place their targets in separate $C_4$ blocks, as explained later.  Since these gates are still linked, the final logical operation has common control lines, as shown in Fig.~\ref{composite_CCZ_overview}.

The second round of error detection uses eight copies of the output of the first round, so 64 $T$~gates are required in total.  The analysis in Sec.~\ref{analysis_section} shows that the error probability for the output state is $3072p^4$ to lowest order.  After decoding the three $C_4$ blocks, the output state is equivalent to the result of applying the composite gate in Fig.~\ref{composite_CCZ_logical} to eight qubits, each of which is in the $\ket{+}$ state.  Section~\ref{teleportation_section} shows how this resource state can teleport the composite-Toffoli gate into any quantum circuit.

\section{Error-detection circuits}
\label{error_detection_section}
The two rounds of error detection in the composite Toffoli gate are (1) building controlled-$S$ gates from $T$~gates and (2) building CCZ gates from controlled-$S$ gates.  The techniques in both rounds are similar, but there are important differences as well.  In this section, we examine the two steps separately for pedagogical clarity.  Furthermore, we assume that Clifford operations are error-free, including initialization and measurement, and that the only errors come from the non-Clifford $T$~gates.

The first round of error detection implements transversal controlled-$H$ gates on a $C_4$ code block.  The first detail we must address is which implementation of the $C_4$ code we use.  All implementations are generated by stabilizers $g_1 = X_1 X_2 X_3 X_4$ and $g_2 = Z_1 Z_2 Z_3 Z_4$, where subscript on each Pauli operator denotes one of the four qubits in the code.  However, logical operators can be chosen in multiple distinct ways, and this choice determines encoding/decoding circuits and the set of transversal gates.  We will label our first implementation the ``$X$/$Y$ encoding'' because the logical $X$ and $Y$ operators on both encoded qubits are weight-2; they can be written as:
\begin{eqnarray}
\overline{X}_1 & = & X_1 X_2 \nonumber \\
\overline{X}_2 & = & X_1 X_3 \nonumber \\
\overline{Y}_1 & = & Y_1 Y_3 \nonumber \\
\overline{Y}_2 & = & Y_1 Y_2,
\label{C4_operators}
\end{eqnarray}
where the bar in $\overline{X}_1$ distinguishes logical code operators from physical qubit operators, and subscript corresponds to one of the two encoded qubits.  Importantly, $X$/$Y$ encoding does not yield a code where transversal CNOT implements encoded CNOT.  The second implementation we use is the standard $X$/$Z$ encoding~\cite{Knill2005}:
\begin{eqnarray}
\overline{X}_1 & = & X_1 X_2 \nonumber \\
\overline{X}_2 & = & X_1 X_3 \nonumber \\
\overline{Z}_1 & = & Z_1 Z_3 \nonumber \\
\overline{Z}_2 & = & Z_1 Z_2.
\label{C4_operators}
\end{eqnarray}
The $X$/$Z$ encoding does permit transversal CNOT, and conversion between encodings will be necessary to satisfy our aim of using only $X$- and $Z$-axis rotations.  Note that our derivation using different encodings is just one way to explain this circuit.  A different interpretation, where there is a single fixed encoding and where all operations that commute with the $C_4$ stabilizers are logical operators, is equally valid.

The $X$/$Y$ encoding permits a transversal $K = T X T^{\dag} = (1/\sqrt{2})(X+Y)$ operation.  In particular, the operator $U = K_1 K_2 K_3 K_4$ commutes with the stabilizers and implements $\overline{K}_1 \overline{K}_2$ and SWAP on the two encoded qubits.  Simply put, $K$ interchanges $X$ and $Y$ operators, just as Hadamard interchanges $X$ and $Z$ operators; in a later step, we map $K$ to $H$.  A similar circuit was used for magic-state distillation with the $C_4$ code using the $X$/$Z$ encoding with transversal $H$~\cite{Meier2012}.

\begin{figure}
  \centering
  \includegraphics[width=8.3cm]{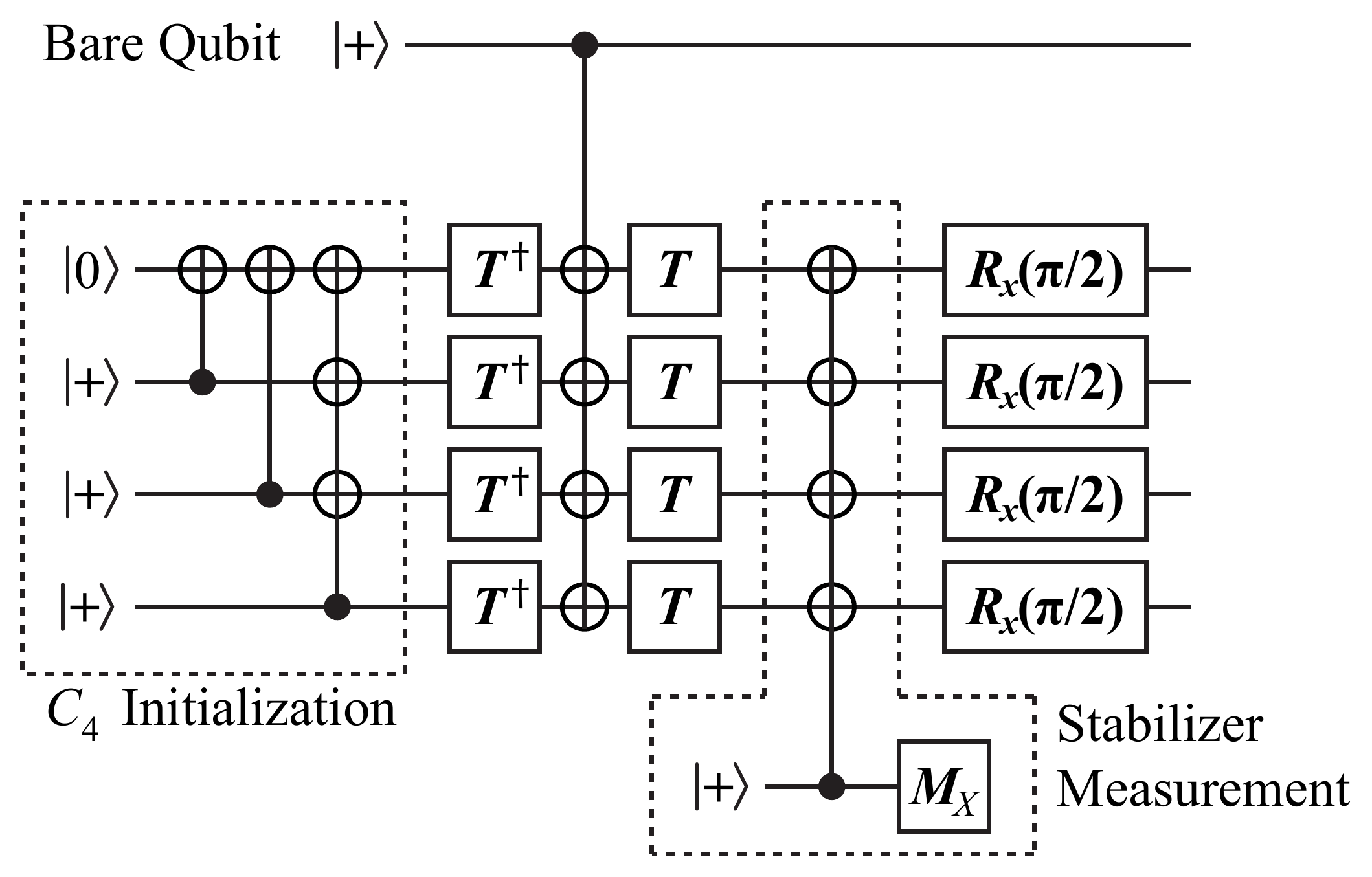}\\
  \caption{Detailed construction of the circuit in Fig.~\ref{C4_controlled_H_simple}.  After initializing in $X$/$Y$ encoding, controlled-$K$ gates are produced using $T$~gates and CNOTs.  The stabilizer measurement can detect a single $Z$ error occurring in any of the $T$~gates.  The transversal $R_x(\pi/2)$ gates transform the $C_4$ block to $X$/$Z$ encoding, and in this basis the controlled-$K$ gates are mapped to controlled-$H$.}
  \label{C4_controlled_H}
\end{figure}

Using $X$/$Y$ encoding, we initialize the circuit to logical $\ket{+}$ qubits, apply controlled-$K$ transversally to the code block, and verify the result.  This procedure is depicted in Fig.~\ref{C4_controlled_H}.  The initialization procedure prepares two encoded $\ket{+}$ qubits as well as a bare $\ket{+}$ qubit.  Next, the $T$~gates and CNOT perform transversal $\overline{K}_1 \overline{K}_2$ and SWAP controlled by the bare qubit.  Since the encoded qubits are identical, the SWAP is trivial.  The stabilizer measurement of the $C_4$ code can detect a single error in any of the $T$~gates.

The final step in this round is to transform this code from $X$/$Y$ to $X$/$Z$ encoding.  The reasons for doing so are twofold.  The $X$/$Z$ encoding has simpler decoding circuits for the $C_4$ block; alternatively, the $X$/$Z$ encoding has transversal, encoded CNOT that enables access to the logical state without decoding.  The code transformation is simple and fault-tolerant.  Apply transversal $R_x(\pi/2) = \exp[i \pi (I - X)/4]$ to each qubit, as shown in Fig.~\ref{C4_controlled_H}.  This operation maps $Y$ operators to $Z$ operators: $[R_x(\pi/2)] Y [R_x(-\pi/2)] = Z$.  The stabilizers are unchanged, but the encoding of logical operators is modified.  $K$ maps to $H$, so the entire circuit is equivalent to applying controlled-$H$ transversally to an $X$/$Z$-encoded $C_4$ block.  The reason for the two-step procedure with $X$/$Y$ and $X$/$Z$ encodings is subtle --- it enables better fault-tolerant circuits because stabilizers can be measured before and after the $R_x(\pi/2)$ gates in Fig.~\ref{C4_controlled_H}.

\begin{figure*}
  \centering
  \includegraphics[width=12cm]{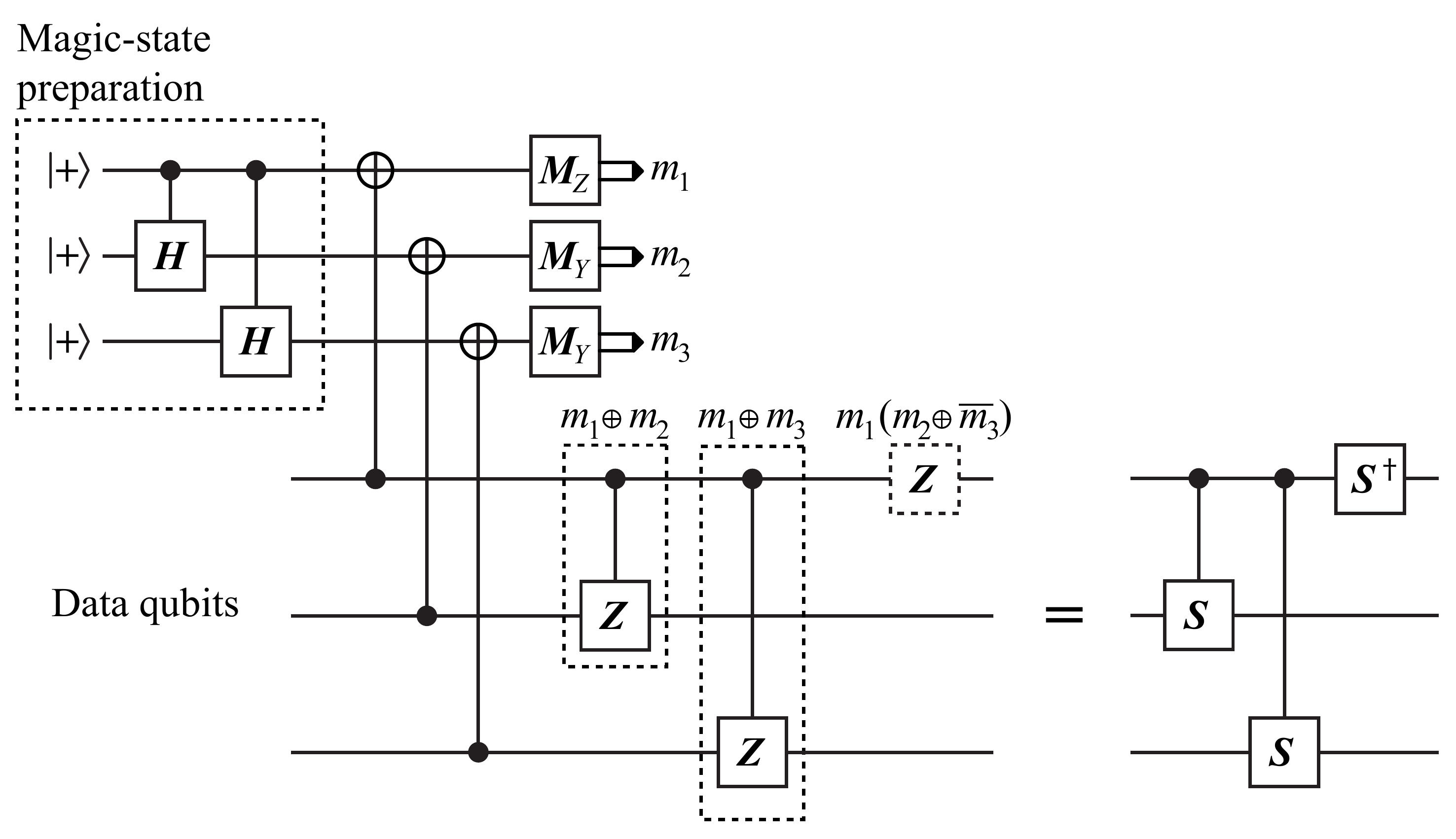}\\
  \caption{Circuit for teleporting two coupled controlled-$S$ gates using the magic state (dashed box in upper left) prepared by the first round of error detection in Fig.~\ref{C4_controlled_H}.  The measurement results are recorded in binary variables $(m_1, m_2, m_3)$.  Subsequent corrections are conditionally implemented based on these measurements, with the conditions for each gate given by the binary expression above the gate.  Overbar here denotes logical inverse, and symbol $\oplus$ denotes binary operation XOR.  The $Z$ operator in the dashed box is incorporated into the Pauli frame~\cite{Knill2005,DiVincenzo2007}.}
  \label{Controlled_H_to_S}
\end{figure*}

The three-qubit magic state created with (effective) controlled-$H$ gates can be used to teleport controlled-$S$ gates.  A circuit for doing so is shown in Fig.~\ref{Controlled_H_to_S}.  The $Y$-basis measurement is not desirable for surface code error correction, but at least one such non-native gate or measurement seems necessary.  We give a fault-tolerant, $C_4$-encoded circuit for this measurement at the end of this section.  The residual $S^{\dag}$ gate will be handled in a later step.

\begin{figure*}
  \centering
  \includegraphics[width=12cm]{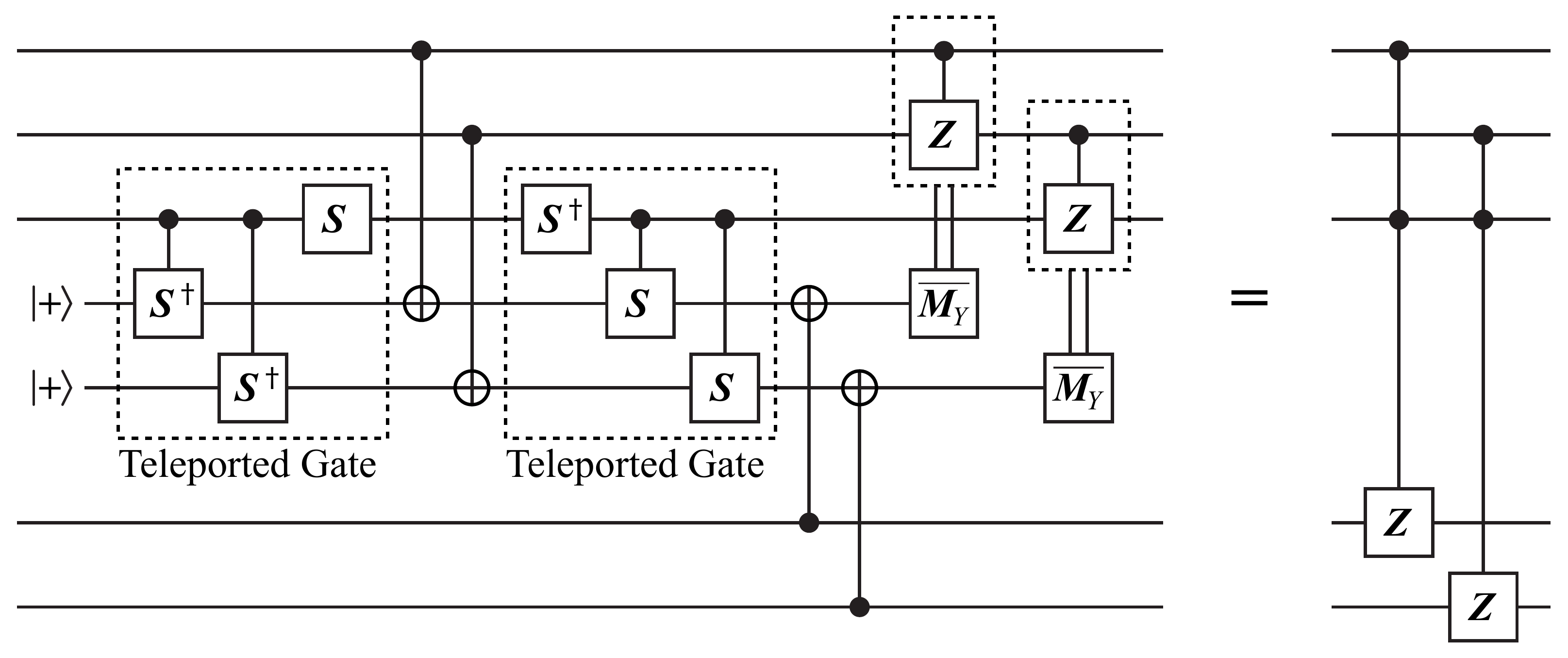}\\
  \caption{Construction of two coupled CCZ gates using teleported controlled-$S$ gates from Fig.~\ref{Controlled_H_to_S}.  The controlled-$S^{\dag}$ gates are created by conceptually applying $Z$ and controlled-$Z$ gates after the output of Fig.~\ref{Controlled_H_to_S}, which in practice is absorbed into the existing conditional operations.  Note that the residual $S$ and $S^{\dag}$ gates cancel.  The overbar in each $\overline{M_y}$ measurement symbol denotes that controlled-$Z$ gates are conditioned on the qubit being in the $(1/\sqrt{2})(\ket{0} - i \ket{1})$ state, the $(-1)$ eigenvector of $Y$.}
  \label{CCZ_construction}
\end{figure*}

The second round of error detection implements transversal controlled-$Z$ between two $C_4$ code blocks, controlled by a bare qubit, as illustrated in Fig.~\ref{composite_CCZ_overview}.  As before, the inputs to the circuit will all be logical $\ket{+}$ qubits.  Controlled-$Z$ is a transversal operation in $C_4$ codes; the logical operation is controlled-$Z$ with swapped targets, which is trivial when the targets are identical.  CCZ gates are broken down into controlled-$S$ gates.  However, the controlled-$S$ magic states from the first round come in coupled pairs which must fan out to separate CCZ gates to ensure that errors in any one $C_4$ block are independent.  The resulting arrangement of CCZ gates with common controls leads to the composite CCZ operation in Fig.~\ref{composite_CCZ_logical}.  A construction for CCZ using controlled-$S$ magic states is shown in Fig.~\ref{CCZ_construction}.  Referring back to Fig.~\ref{composite_CCZ_overview}, we see that each of the four adjacent pairs of coupled-CCZ gates (sharing one common control) is implemented by the circuit in Fig.~\ref{CCZ_construction}.  Each coupled-CCZ gate uses two copies of Fig.~\ref{C4_controlled_H}, or 16 $T$~gates.  The entire circuit thus uses 64 $T$~gates.

Although Fig.~\ref{CCZ_construction} builds a coupled pair of CCZ gates, each has distance two with respect to $T$-gate error with probability $p$, resulting in total error probability $56p^2$ to leading order (using error detection in the first round only).  By using another round of error detection with $C_4$ codes, we can achieve distance four and error probability of $3072p^4$ for a composite operation of four CCZ gates.  The next section calculates error probability of this composite CCZ gate when one assumes that $T$~gates are the dominant failure mechanism.

\begin{figure}
  \centering
  \includegraphics[width=8.3cm]{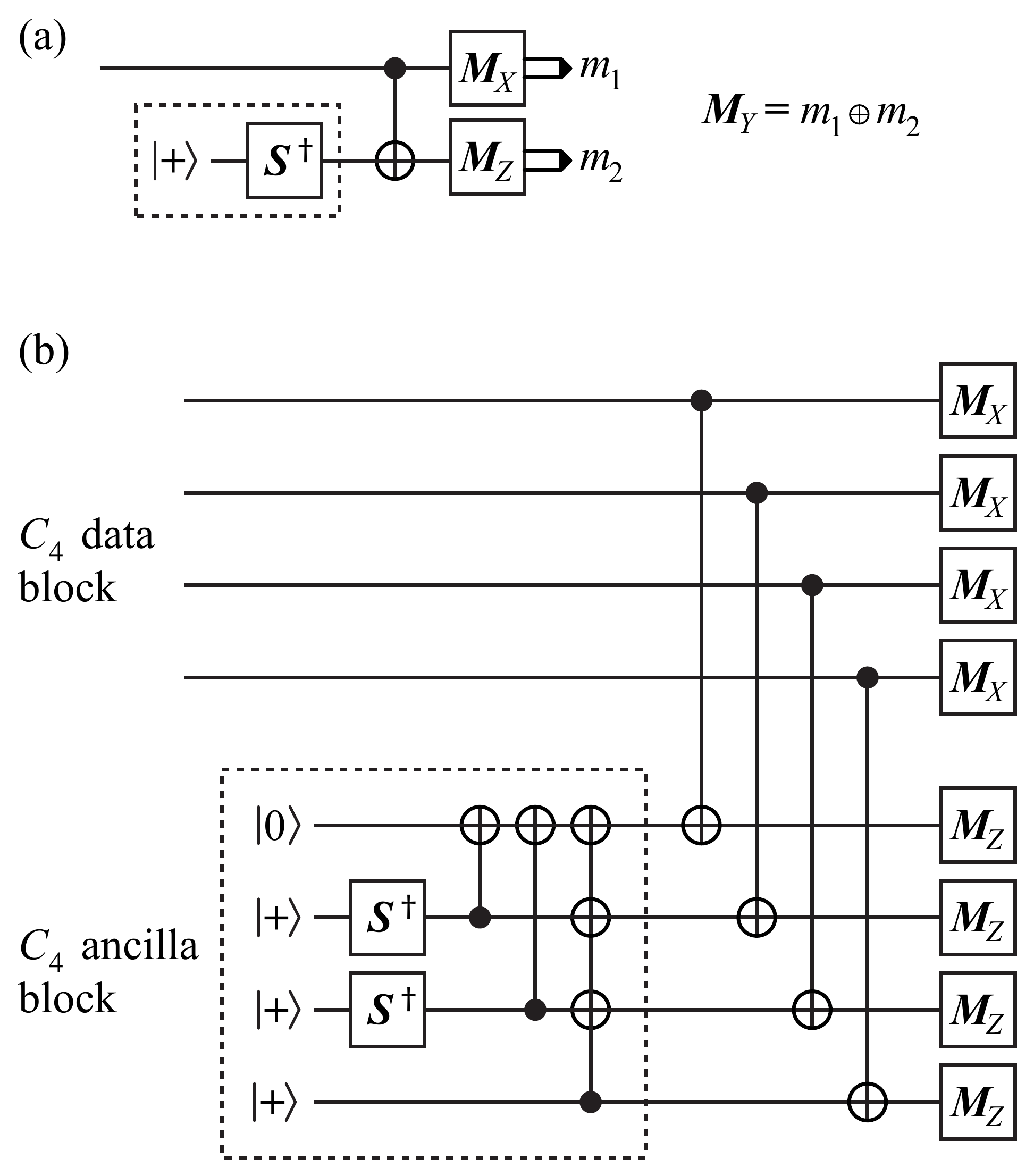}\\
  \caption{$Y$-basis measurement using ancillas and operations transversal in $C_4$ codes.  \textbf{(a)} Logical measurement circuit using the $S^{\dag}\ket{+}$ ancilla state.  The measurement result is $M_y = m_1 \oplus m_2$, where $\oplus$ denotes binary XOR. \textbf{(b)} $Y$-basis measurement in $C_4$ code blocks.  The dashed box shows an encoding circuit for the ancilla block that is prepared, though distillation of this register may also be required.  Both logical $Y$-basis measurement and code stabilizers can be reconstructed from the single-qubit measurements.}
  \label{Y_measurement_circuit}
\end{figure}

The final circuit component we require is a fault-tolerant $Y$-basis measurement $M_y$.  A simple way to do this is to perform the gate $R_x(\pi/2)$ followed by $Z$-basis measurement.  However, our circuit constructions use $C_4$-encoded qubits, so we would like to perform $C_4$-encoded $M_y$.  The logical operation $R_x(\pi/2)$ is not transversal in $C_4$, so it is not convenient to implement.  However, we can implement $M_y$ using operations transversal in $C_4$ with the aid of the ancilla state $S^{\dag}\ket{+}$, as shown in Fig.~\ref{Y_measurement_circuit}a.  The $Y$-basis measurement is given by the binary XOR of the $M_x$ and $M_z$ results.  By encoding two $S^{\dag}\ket{+}$ qubits in a $C_4$ code, we can perform encoded $M_y$ using transversal operations, as shown in Fig~\ref{Y_measurement_circuit}b.  This is fault-tolerant $M_y$ with respect to the $C_4$ code blocks, because the single-qubit measurements can be used to reconstruct both the logical $Y$-basis measurements and the stabilizer parity measurements for error detection.  State $S^{\dag}\ket{+}$ is not natively available in the surface code, so it may require distillation~\cite{Raussendorf2007,Fowler2009,Fowler2012d}.  The protocol in Ref.~\cite{Aliferis_thesis} (p. 94) can be adapted to distilling $C_4$-encoded $S^{\dag}\ket{+}$ qubits.

\section{Error analysis}
\label{analysis_section}
Determining the probability of error in the output of the composite-Toffoli circuit is simplified by the operating assumption that errors only occur in $T$~gates with independent probability $p$.  We assume that $p \ll 1$ so that the output error is approximated well by the first non-vanishing term in a power-series expansion in $p$, and we show that this term is $O(p^4)$.  As before, we analyze the two rounds of the protocol, where the second round detects some errors missed in the first.

In the first round of error detection shown in Fig.~\ref{C4_controlled_H}, there are eight $T$~gates which may each have a $Z$~error.  Any single error will be detected by the $C_4$ stabilizers, while any combination of two errors will not be detected.  There are 28 distinct arrangements of two errors, and they can be grouped into seven error patterns at the output.  After teleportation of coupled-controlled-$S$ gates in Fig.~\ref{Controlled_H_to_S}, the possible error configurations are the seven configurations of one or more $Z$~errors on the three output qubits.  Each of these configurations has probability $4p^2$ because each can arise in four different patterns of $T$-gate errors.  The total error probability for this operation is $28p^2$, as expected.

An important design feature of the composite Toffoli gate is that the three outputs of the first round fan out to different $C_4$ blocks in the second round.  The most likely patterns of errors which evade detection in the second round are those where two instances of the first round both had undetected errors at their respective outputs.  As before, a single error in any $C_4$ block will be detected, so the two faulty instances of coupled-controlled-$S$ must have exactly the same error configuration.  If not, there will be a single error in at least one block, which is detected.  None of the $C_4$ codes detect errors when any two first-round states have matched errors, so these events represent the most likely errors at the output of the second round.  For a few configurations, errors from the first round can cancel.  Referring to Fig.~\ref{CCZ_construction}, if both coupled-controlled-$S$ instances have a single $Z$ error on the $S$/$S^{\dag}$ qubit, these will cancel without any effect on the broader circuit.  There are four different possible patterns for this event.  As a result, undetected output errors can occur via six matching first-round error patterns, each having 28 permutations, or the seventh first-round pattern with just 24 permutations (the other four self-cancel), which adds up to 192 distinct configurations.  Each first-round error pattern has probability $4p^2$, so the total probability of error in the composite CCZ gate is $192 \times (4p^2)^2 = 3072p^4$.

The use of error detection, instead of correction, implies that known faulty states are discarded.  In such an event, some or all of the preparation steps must be repeated.  The probability of detected circuit failure can be upper bounded by $p_{\mathrm{fail}} \le 1 - (1-p)^{64} \le 64p$.  This assumes the entire circuit fails on any single $T$-gate error.  Less overhead from repeating circuits is required if one repeats only the round which failed; if one of the eight copies of first-round error detection fails, repeat just that circuit rather than the entire composite CCZ gate.  To accommodate failure, we prepare encoded states before teleporting data through the gate.

\section{Teleportation into quantum algorithms}
\label{teleportation_section}
\begin{figure*}
  \centering
  \includegraphics[width=\textwidth]{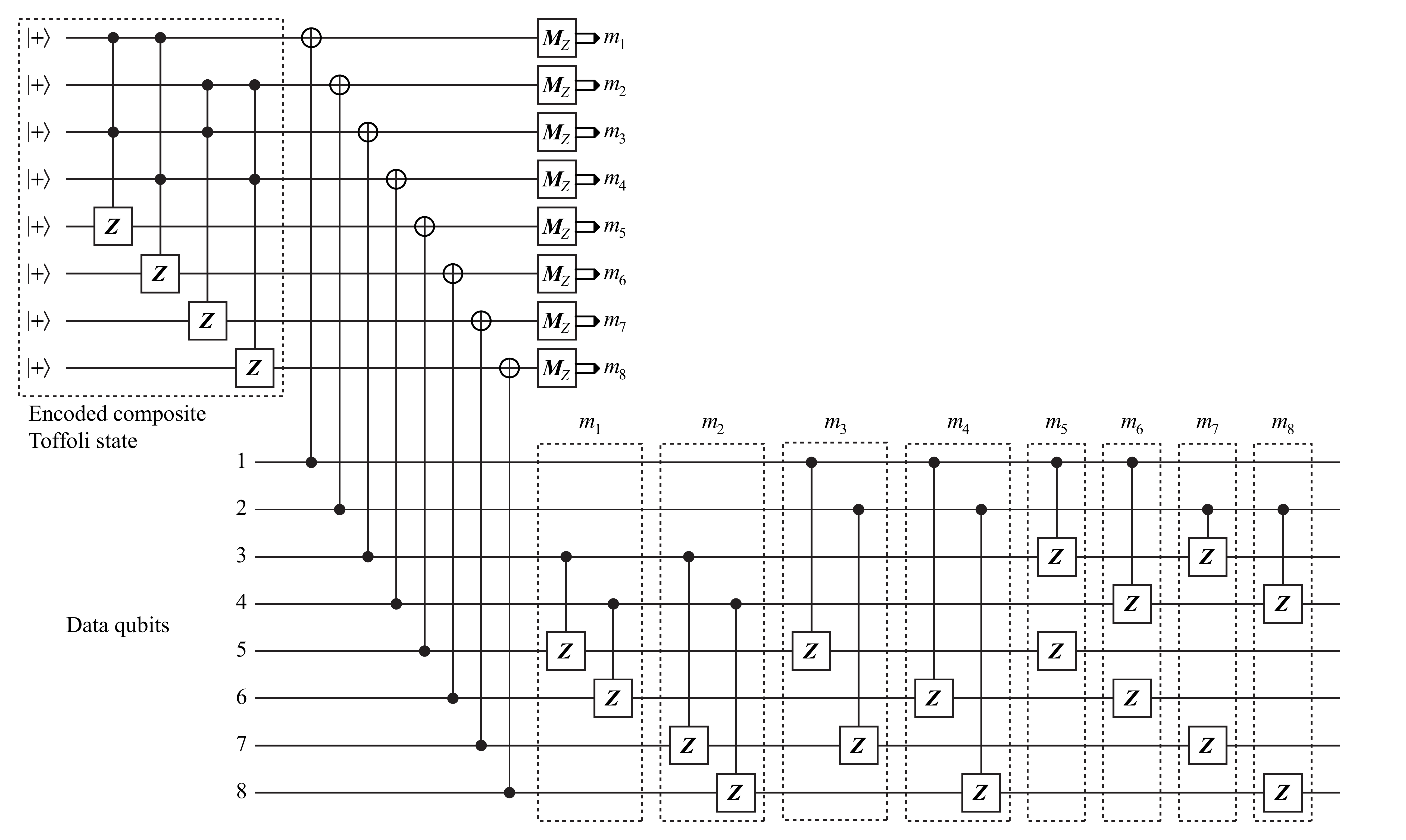}\\
  \caption{Teleportation circuit for the composite CCZ gate shown in Fig.~\ref{composite_CCZ_logical}.  CCZ~gates are symmetric in their inputs, so placing Hadamard gates on both sides of the teleportation circuit on the same data qubit will convert the affected CCZ gates to Toffoli (targeting this same qubit).  Operations in dashed boxes, which are all in the Clifford group, are implemented conditioned on the indicated measurement result being logical $\ket{1}$.  In many cases, the conditional corrections can be delayed or combined with other gates.}
  \label{composite_CCZ_teleportation}
\end{figure*}

The composite CCZ operation (or equivalently composite Toffoli) in Fig.~\ref{composite_CCZ_logical} can be encoded into a quantum register by applying this gate to eight $\ket{+}$ qubits.  After constructing and verifying this state, the gate interacts with data qubits using teleportation, which is an extension of the methods developed in Ref.~\cite{Gottesman1999}.  The teleportation circuit is shown in Fig.~\ref{composite_CCZ_teleportation}.

If the four coupled CCZ operations are problematic, one can sacrifice two CCZ gates to leave two uncoupled CCZ gates.  Referring to Fig.~\ref{composite_CCZ_logical}, if one sets inputs 6 and 7 to $\ket{0}$ while the others are set to $\ket{+}$, then the second and third CCZ gates act trivially.  Equivalently, the teleportation circuit in Fig.~\ref{composite_CCZ_teleportation} is modified by deleting lines 6 and 7, as well as any gates which touch them.  As a result, two independent CCZ gates are produced.  The total error probability will be lower because some errors become trivial.

\section{Discussion}
\label{discussion_section}
Using our results, circuits that depend on Toffoli gates have reduced fault-tolerant resource overhead. The precise improvement factor depends generally on too many parameters and assumptions to be covered here.  Instead, we give an illustrative example showing how resource costs are lowered.  Suppose that we are using surface code error correction as in Refs.~\cite{Jones2012,Fowler2012b}.  A typical implementation of Shor's algorithm~\cite{Shor1999} may require an error probability per Toffoli gate around $10^{-12}$.  The simplest Toffoli circuit uses four $T$~gates, which would each require error probability $2 \times 10^{-13}$.  The error-detecting constructions in Refs.~\cite{Jones2013,Eastin2012} require eight $T$~gates with error probability $2 \times 10^{-7}$.  Increasing the acceptable probability of error means one less round of magic-state distillation is required, reducing total resources by about a factor of ten~\cite{Fowler2012b,Jones2013}.  Additionally, intermediate Clifford operations can use lower code distance~\cite{Fowler2012b}.  The construction in this paper continues this trend.  With two rounds of error detection, the $T$-gate error probability need only be $10^{-4}$.  Even less magic-state distillation is required in this instance, and intermediate operations can tolerate higher probability of errors.  By a cursory resource counting, the savings can be a factor of 20 to 50 for producing Toffoli gates, using methods developed in Refs.~\cite{Fowler2012b,Jones2013}.  Moreover, error rate $10^{-4}$ is plausibly achievable by physical gates without error correction, which could make magic-state distillation unnecessary and save even more resources.

Some important considerations must be mentioned.  We assumed that non-Clifford operations dominate resource costs, which was borne out in previous investigations~\cite{Isailovic2008,Jones2012,Fowler2012b,Fowler2012d,Fowler2013}.  However, the composite Toffoli changes the situation when its resource cost no longer dominates the the total cost of the computation.  Other operations in a quantum algorithm like routing of qubits for long-range interactions may become important.  The resource savings factor for the entire algorithm will always be less than that for the individual Toffoli gates; still, most quantum algorithms like factoring~\cite{Shor1999,Isailovic2008,Jones2012} and simulation~\cite{Lloyd1996,Jones2012b} benefit substantially from a more efficient Toffoli construction.

More research is needed to fully understand resource costs of the composite Toffoli construction in the context of a chosen quantum code.  Similar work has been performed to optimize magic-state distillation protocols implemented in a surface code~\cite{Fowler2012d,Fowler2013}.  Our results could also be implemented within other codes, such as Bacon-Shor codes~\cite{Aliferis2007}.  In such analysis, another opportunity beyond $T$~gates for saving resources is in the Clifford gates.  We have assumed throughout that Clifford operations are perfect, but this is never the case in practice.  Instead, Clifford operations can have arbitrarily low error for some resource cost.  The constructions in this paper use $C_4$ codes to detect errors in $T$~gates, but they can also detect errors in other gates~\cite{Knill2005}.  For example, the Clifford operations produced using a surface code could have higher error rate if one knew that errors would be caught by the $C_4$ error-detection circuits.  When higher error rates are allowed, lower code distance can be used, which means fewer hardware resources are required for the same circuit.

The composite Toffoli gate demonstrates several important techniques in fault-tolerant quantum computing that merit further investigation.  A quantum operation is encoded into a known state that is verified before being teleported into the rest of the quantum circuit.  Early work on teleportation gates focused on one-, two-, or three-qubit operations~\cite{Gottesman1999,Nielsen2000}; by comparison, the composite Toffoli gate is an eight-qubit operation.  The process of compiling quantum operations into encoded states with verification followed by teleportation is a powerful technique for generating fault-tolerant quantum circuits.  We propose the term \emph{quantum logic synthesis} for methods of synthesizing arbitrary-size, fault-tolerant quantum logic networks in a hierarchical arrangement of preparation and teleportation.  The possible techniques go far beyond ``sequential'' decompositions~\cite{Isailovic2008,Jones2012,Jones2012b,Fowler2012b}, where a quantum algorithm is decomposed into a long sequence of fundamental gates from a small set.  For each fundamental gate, fault-tolerant constructions are known, but the cost of each is high because every operation must have very low error rate.  By contrast, hierarchical designs weave error checking into the algorithm, allowing higher error rates throughout.  Quantum logic synthesis can compress larger, more complex operations than Toffoli gates, which is the subject of forthcoming work.


\begin{thebibliography}{37}%
\makeatletter
\providecommand \@ifxundefined [1]{%
 \@ifx{#1\undefined}
}%
\providecommand \@ifnum [1]{%
 \ifnum #1\expandafter \@firstoftwo
 \else \expandafter \@secondoftwo
 \fi
}%
\providecommand \@ifx [1]{%
 \ifx #1\expandafter \@firstoftwo
 \else \expandafter \@secondoftwo
 \fi
}%
\providecommand \natexlab [1]{#1}%
\providecommand \enquote  [1]{``#1''}%
\providecommand \bibnamefont  [1]{#1}%
\providecommand \bibfnamefont [1]{#1}%
\providecommand \citenamefont [1]{#1}%
\providecommand \href@noop [0]{\@secondoftwo}%
\providecommand \href [0]{\begingroup \@sanitize@url \@href}%
\providecommand \@href[1]{\@@startlink{#1}\@@href}%
\providecommand \@@href[1]{\endgroup#1\@@endlink}%
\providecommand \@sanitize@url [0]{\catcode `\\12\catcode `\$12\catcode
  `\&12\catcode `\#12\catcode `\^12\catcode `\_12\catcode `\%12\relax}%
\providecommand \@@startlink[1]{}%
\providecommand \@@endlink[0]{}%
\providecommand \url  [0]{\begingroup\@sanitize@url \@url }%
\providecommand \@url [1]{\endgroup\@href {#1}{\urlprefix }}%
\providecommand \urlprefix  [0]{URL }%
\providecommand \Eprint [0]{\href }%
\providecommand \doibase [0]{http://dx.doi.org/}%
\providecommand \selectlanguage [0]{\@gobble}%
\providecommand \bibinfo  [0]{\@secondoftwo}%
\providecommand \bibfield  [0]{\@secondoftwo}%
\providecommand \translation [1]{[#1]}%
\providecommand \BibitemOpen [0]{}%
\providecommand \bibitemStop [0]{}%
\providecommand \bibitemNoStop [0]{.\EOS\space}%
\providecommand \EOS [0]{\spacefactor3000\relax}%
\providecommand \BibitemShut  [1]{\csname bibitem#1\endcsname}%
\let\auto@bib@innerbib\@empty
\bibitem [{\citenamefont {Calderbank}\ and\ \citenamefont
  {Shor}(1996)}]{Calderbank1996}%
  \BibitemOpen
  \bibfield  {author} {\bibinfo {author} {\bibfnamefont {A.~R.}\ \bibnamefont
  {Calderbank}}\ and\ \bibinfo {author} {\bibfnamefont {P.~W.}\ \bibnamefont
  {Shor}},\ }\href {\doibase 10.1103/PhysRevA.54.1098} {\bibfield  {journal}
  {\bibinfo  {journal} {Phys. Rev. A}\ }\textbf {\bibinfo {volume} {54}},\
  \bibinfo {pages} {1098} (\bibinfo {year} {1996})}\BibitemShut {NoStop}%
\bibitem [{\citenamefont {Steane}(1996)}]{Steane1996}%
  \BibitemOpen
  \bibfield  {author} {\bibinfo {author} {\bibfnamefont {A.}~\bibnamefont
  {Steane}},\ }\href {\doibase 10.1098/rspa.1996.0136} {\bibfield  {journal}
  {\bibinfo  {journal} {Proceedings of the Royal Society of London. Series A:
  Mathematical, Physical and Engineering Sciences}\ }\textbf {\bibinfo {volume}
  {452}},\ \bibinfo {pages} {2551} (\bibinfo {year} {1996})}\BibitemShut
  {NoStop}%
\bibitem [{\citenamefont {Preskill}(1998)}]{Preskill1998}%
  \BibitemOpen
  \bibfield  {author} {\bibinfo {author} {\bibfnamefont {J.}~\bibnamefont
  {Preskill}},\ }\href {\doibase 10.1098/rspa.1998.0167} {\bibfield  {journal}
  {\bibinfo  {journal} {Proceedings of the Royal Society of London. Series A:
  Mathematical, Physical and Engineering Sciences}\ }\textbf {\bibinfo {volume}
  {454}},\ \bibinfo {pages} {385} (\bibinfo {year} {1998})}\BibitemShut
  {NoStop}%
\bibitem [{\citenamefont {Nielsen}\ and\ \citenamefont
  {Chuang}(2000)}]{Nielsen2000}%
  \BibitemOpen
  \bibfield  {author} {\bibinfo {author} {\bibfnamefont {M.~A.}\ \bibnamefont
  {Nielsen}}\ and\ \bibinfo {author} {\bibfnamefont {I.~L.}\ \bibnamefont
  {Chuang}},\ }\href@noop {} {\emph {\bibinfo {title} {Quantum Computation and
  Quantum Information}}},\ \bibinfo {edition} {1st}\ ed.\ (\bibinfo
  {publisher} {Cambridge University Press},\ \bibinfo {year}
  {2000})\BibitemShut {NoStop}%
\bibitem [{\citenamefont {Knill}(2005)}]{Knill2005}%
  \BibitemOpen
  \bibfield  {author} {\bibinfo {author} {\bibfnamefont {E.}~\bibnamefont
  {Knill}},\ }\href@noop {} {\bibfield  {journal} {\bibinfo  {journal}
  {Nature}\ }\textbf {\bibinfo {volume} {434}},\ \bibinfo {pages} {39}
  (\bibinfo {year} {2005})}\BibitemShut {NoStop}%
\bibitem [{\citenamefont {Isailovic}\ \emph {et~al.}(2008)\citenamefont
  {Isailovic}, \citenamefont {Whitney}, \citenamefont {Patel},\ and\
  \citenamefont {Kubiatowicz}}]{Isailovic2008}%
  \BibitemOpen
  \bibfield  {author} {\bibinfo {author} {\bibfnamefont {N.}~\bibnamefont
  {Isailovic}}, \bibinfo {author} {\bibfnamefont {M.}~\bibnamefont {Whitney}},
  \bibinfo {author} {\bibfnamefont {Y.}~\bibnamefont {Patel}}, \ and\ \bibinfo
  {author} {\bibfnamefont {J.}~\bibnamefont {Kubiatowicz}},\ }in\ \href@noop {}
  {\emph {\bibinfo {booktitle} {35th International Symposium on Computer
  Architecture, 2008 (ISCA'08)}}}\ (\bibinfo {year} {2008})\BibitemShut
  {NoStop}%
\bibitem [{\citenamefont {Jones}\ \emph
  {et~al.}(2012{\natexlab{a}})\citenamefont {Jones}, \citenamefont {{Van
  Meter}}, \citenamefont {Fowler}, \citenamefont {McMahon}, \citenamefont
  {Kim}, \citenamefont {Ladd},\ and\ \citenamefont {Yamamoto}}]{Jones2012}%
  \BibitemOpen
  \bibfield  {author} {\bibinfo {author} {\bibfnamefont {N.~C.}\ \bibnamefont
  {Jones}}, \bibinfo {author} {\bibfnamefont {R.}~\bibnamefont {{Van Meter}}},
  \bibinfo {author} {\bibfnamefont {A.~G.}\ \bibnamefont {Fowler}}, \bibinfo
  {author} {\bibfnamefont {P.~L.}\ \bibnamefont {McMahon}}, \bibinfo {author}
  {\bibfnamefont {J.}~\bibnamefont {Kim}}, \bibinfo {author} {\bibfnamefont
  {T.~D.}\ \bibnamefont {Ladd}}, \ and\ \bibinfo {author} {\bibfnamefont
  {Y.}~\bibnamefont {Yamamoto}},\ }\href@noop {} {\bibfield  {journal}
  {\bibinfo  {journal} {Phys. Rev. X}\ }\textbf {\bibinfo {volume} {2}},\
  \bibinfo {pages} {031007} (\bibinfo {year} {2012}{\natexlab{a}})}\BibitemShut
  {NoStop}%
\bibitem [{\citenamefont {Fowler}\ \emph {et~al.}(2012)\citenamefont {Fowler},
  \citenamefont {Mariantoni}, \citenamefont {Martinis},\ and\ \citenamefont
  {Cleland}}]{Fowler2012b}%
  \BibitemOpen
  \bibfield  {author} {\bibinfo {author} {\bibfnamefont {A.~G.}\ \bibnamefont
  {Fowler}}, \bibinfo {author} {\bibfnamefont {M.}~\bibnamefont {Mariantoni}},
  \bibinfo {author} {\bibfnamefont {J.~M.}\ \bibnamefont {Martinis}}, \ and\
  \bibinfo {author} {\bibfnamefont {A.~N.}\ \bibnamefont {Cleland}},\ }\href
  {\doibase 10.1103/PhysRevA.86.032324} {\bibfield  {journal} {\bibinfo
  {journal} {Phys. Rev. A}\ }\textbf {\bibinfo {volume} {86}},\ \bibinfo
  {pages} {032324} (\bibinfo {year} {2012})}\BibitemShut {NoStop}%
\bibitem [{\citenamefont {Ladd}\ \emph {et~al.}(2010)\citenamefont {Ladd},
  \citenamefont {Jelezko}, \citenamefont {Laflamme}, \citenamefont {Nakamura},
  \citenamefont {Monroe},\ and\ \citenamefont {{O'Brien}}}]{Ladd2010}%
  \BibitemOpen
  \bibfield  {author} {\bibinfo {author} {\bibfnamefont {T.~D.}\ \bibnamefont
  {Ladd}}, \bibinfo {author} {\bibfnamefont {F.}~\bibnamefont {Jelezko}},
  \bibinfo {author} {\bibfnamefont {R.}~\bibnamefont {Laflamme}}, \bibinfo
  {author} {\bibfnamefont {Y.}~\bibnamefont {Nakamura}}, \bibinfo {author}
  {\bibfnamefont {C.}~\bibnamefont {Monroe}}, \ and\ \bibinfo {author}
  {\bibfnamefont {J.~L.}\ \bibnamefont {{O'Brien}}},\ }\href@noop {} {\bibfield
   {journal} {\bibinfo  {journal} {Nature}\ }\textbf {\bibinfo {volume}
  {464}},\ \bibinfo {pages} {45} (\bibinfo {year} {2010})}\BibitemShut
  {NoStop}%
\bibitem [{\citenamefont {Lucero}\ \emph {et~al.}(2012)\citenamefont {Lucero},
  \citenamefont {Barends}, \citenamefont {Chen}, \citenamefont {Kelly},
  \citenamefont {Mariantoni}, \citenamefont {Megrant}, \citenamefont
  {{O'Malley}}, \citenamefont {Sank}, \citenamefont {Vainsencher},
  \citenamefont {Wenner}, \citenamefont {White}, \citenamefont {Yin},
  \citenamefont {Cleland},\ and\ \citenamefont {Martinis}}]{Lucero2012}%
  \BibitemOpen
  \bibfield  {author} {\bibinfo {author} {\bibfnamefont {E.}~\bibnamefont
  {Lucero}}, \bibinfo {author} {\bibfnamefont {R.}~\bibnamefont {Barends}},
  \bibinfo {author} {\bibfnamefont {Y.}~\bibnamefont {Chen}}, \bibinfo {author}
  {\bibfnamefont {J.}~\bibnamefont {Kelly}}, \bibinfo {author} {\bibfnamefont
  {M.}~\bibnamefont {Mariantoni}}, \bibinfo {author} {\bibfnamefont
  {A.}~\bibnamefont {Megrant}}, \bibinfo {author} {\bibfnamefont
  {P.}~\bibnamefont {{O'Malley}}}, \bibinfo {author} {\bibfnamefont
  {D.}~\bibnamefont {Sank}}, \bibinfo {author} {\bibfnamefont {A.}~\bibnamefont
  {Vainsencher}}, \bibinfo {author} {\bibfnamefont {J.}~\bibnamefont {Wenner}},
  \bibinfo {author} {\bibfnamefont {T.}~\bibnamefont {White}}, \bibinfo
  {author} {\bibfnamefont {Y.}~\bibnamefont {Yin}}, \bibinfo {author}
  {\bibfnamefont {A.~N.}\ \bibnamefont {Cleland}}, \ and\ \bibinfo {author}
  {\bibfnamefont {J.~M.}\ \bibnamefont {Martinis}},\ }\href@noop {} {\bibfield
  {journal} {\bibinfo  {journal} {Nature Physics}\ }\textbf {\bibinfo {volume}
  {8}},\ \bibinfo {pages} {719} (\bibinfo {year} {2012})}\BibitemShut {NoStop}%
\bibitem [{\citenamefont {Blatt}\ and\ \citenamefont {Roos}(2012)}]{Blatt2012}%
  \BibitemOpen
  \bibfield  {author} {\bibinfo {author} {\bibfnamefont {R.}~\bibnamefont
  {Blatt}}\ and\ \bibinfo {author} {\bibfnamefont {C.~F.}\ \bibnamefont
  {Roos}},\ }\href@noop {} {\bibfield  {journal} {\bibinfo  {journal} {Nature
  Physics}\ }\textbf {\bibinfo {volume} {8}},\ \bibinfo {pages} {277} (\bibinfo
  {year} {2012})}\BibitemShut {NoStop}%
\bibitem [{\citenamefont {Politi}\ \emph {et~al.}(2009)\citenamefont {Politi},
  \citenamefont {Matthews},\ and\ \citenamefont {O'Brien}}]{Politi2009}%
  \BibitemOpen
  \bibfield  {author} {\bibinfo {author} {\bibfnamefont {A.}~\bibnamefont
  {Politi}}, \bibinfo {author} {\bibfnamefont {J.~C.~F.}\ \bibnamefont
  {Matthews}}, \ and\ \bibinfo {author} {\bibfnamefont {J.~L.}\ \bibnamefont
  {O'Brien}},\ }\href {\doibase 10.1126/science.1173731} {\bibfield  {journal}
  {\bibinfo  {journal} {Science}\ }\textbf {\bibinfo {volume} {325}},\ \bibinfo
  {pages} {1221} (\bibinfo {year} {2009})}\BibitemShut {NoStop}%
\bibitem [{\citenamefont {Maurer}\ \emph {et~al.}(2012)\citenamefont {Maurer},
  \citenamefont {Kucsko}, \citenamefont {Latta}, \citenamefont {Jiang},
  \citenamefont {Yao}, \citenamefont {Bennett}, \citenamefont {Pastawski},
  \citenamefont {Hunger}, \citenamefont {Chisholm}, \citenamefont {Markham},
  \citenamefont {Twitchen}, \citenamefont {Cirac},\ and\ \citenamefont
  {Lukin}}]{Maurer2012}%
  \BibitemOpen
  \bibfield  {author} {\bibinfo {author} {\bibfnamefont {P.~C.}\ \bibnamefont
  {Maurer}}, \bibinfo {author} {\bibfnamefont {G.}~\bibnamefont {Kucsko}},
  \bibinfo {author} {\bibfnamefont {C.}~\bibnamefont {Latta}}, \bibinfo
  {author} {\bibfnamefont {L.}~\bibnamefont {Jiang}}, \bibinfo {author}
  {\bibfnamefont {N.~Y.}\ \bibnamefont {Yao}}, \bibinfo {author} {\bibfnamefont
  {S.~D.}\ \bibnamefont {Bennett}}, \bibinfo {author} {\bibfnamefont
  {F.}~\bibnamefont {Pastawski}}, \bibinfo {author} {\bibfnamefont
  {D.}~\bibnamefont {Hunger}}, \bibinfo {author} {\bibfnamefont
  {N.}~\bibnamefont {Chisholm}}, \bibinfo {author} {\bibfnamefont
  {M.}~\bibnamefont {Markham}}, \bibinfo {author} {\bibfnamefont {D.~J.}\
  \bibnamefont {Twitchen}}, \bibinfo {author} {\bibfnamefont {J.~I.}\
  \bibnamefont {Cirac}}, \ and\ \bibinfo {author} {\bibfnamefont {M.~D.}\
  \bibnamefont {Lukin}},\ }\href {\doibase 10.1126/science.1220513} {\bibfield
  {journal} {\bibinfo  {journal} {Science}\ }\textbf {\bibinfo {volume}
  {336}},\ \bibinfo {pages} {1283} (\bibinfo {year} {2012})}\BibitemShut
  {NoStop}%
\bibitem [{\citenamefont {Shulman}\ \emph {et~al.}(2012)\citenamefont
  {Shulman}, \citenamefont {Dial}, \citenamefont {Harvey}, \citenamefont
  {Bluhm}, \citenamefont {Umansky},\ and\ \citenamefont
  {Yacoby}}]{Shulman2012}%
  \BibitemOpen
  \bibfield  {author} {\bibinfo {author} {\bibfnamefont {M.~D.}\ \bibnamefont
  {Shulman}}, \bibinfo {author} {\bibfnamefont {O.~E.}\ \bibnamefont {Dial}},
  \bibinfo {author} {\bibfnamefont {S.~P.}\ \bibnamefont {Harvey}}, \bibinfo
  {author} {\bibfnamefont {H.}~\bibnamefont {Bluhm}}, \bibinfo {author}
  {\bibfnamefont {V.}~\bibnamefont {Umansky}}, \ and\ \bibinfo {author}
  {\bibfnamefont {A.}~\bibnamefont {Yacoby}},\ }\href {\doibase
  10.1126/science.1217692} {\bibfield  {journal} {\bibinfo  {journal}
  {Science}\ }\textbf {\bibinfo {volume} {336}},\ \bibinfo {pages} {202}
  (\bibinfo {year} {2012})}\BibitemShut {NoStop}%
\bibitem [{\citenamefont {Zeng}\ \emph {et~al.}(2011)\citenamefont {Zeng},
  \citenamefont {Cross},\ and\ \citenamefont {Chuang}}]{Zeng2007}%
  \BibitemOpen
  \bibfield  {author} {\bibinfo {author} {\bibfnamefont {B.}~\bibnamefont
  {Zeng}}, \bibinfo {author} {\bibfnamefont {A.}~\bibnamefont {Cross}}, \ and\
  \bibinfo {author} {\bibfnamefont {I.}~\bibnamefont {Chuang}},\ }\href
  {\doibase 10.1109/TIT.2011.2161917} {\bibfield  {journal} {\bibinfo
  {journal} {Information Theory, IEEE Transactions on}\ }\textbf {\bibinfo
  {volume} {57}},\ \bibinfo {pages} {6272} (\bibinfo {year}
  {2011})}\BibitemShut {NoStop}%
\bibitem [{\citenamefont {Eastin}\ and\ \citenamefont
  {Knill}(2009)}]{Eastin2009}%
  \BibitemOpen
  \bibfield  {author} {\bibinfo {author} {\bibfnamefont {B.}~\bibnamefont
  {Eastin}}\ and\ \bibinfo {author} {\bibfnamefont {E.}~\bibnamefont {Knill}},\
  }\href {\doibase 10.1103/PhysRevLett.102.110502} {\bibfield  {journal}
  {\bibinfo  {journal} {Phys. Rev. Lett.}\ }\textbf {\bibinfo {volume} {102}},\
  \bibinfo {pages} {110502} (\bibinfo {year} {2009})}\BibitemShut {NoStop}%
\bibitem [{\citenamefont {Barenco}\ \emph {et~al.}(1995)\citenamefont
  {Barenco}, \citenamefont {Bennett}, \citenamefont {Cleve}, \citenamefont
  {DiVincenzo}, \citenamefont {Margolus}, \citenamefont {Shor}, \citenamefont
  {Sleator}, \citenamefont {Smolin},\ and\ \citenamefont
  {Weinfurter}}]{Barenco1995}%
  \BibitemOpen
  \bibfield  {author} {\bibinfo {author} {\bibfnamefont {A.}~\bibnamefont
  {Barenco}}, \bibinfo {author} {\bibfnamefont {C.~H.}\ \bibnamefont
  {Bennett}}, \bibinfo {author} {\bibfnamefont {R.}~\bibnamefont {Cleve}},
  \bibinfo {author} {\bibfnamefont {D.~P.}\ \bibnamefont {DiVincenzo}},
  \bibinfo {author} {\bibfnamefont {N.}~\bibnamefont {Margolus}}, \bibinfo
  {author} {\bibfnamefont {P.}~\bibnamefont {Shor}}, \bibinfo {author}
  {\bibfnamefont {T.}~\bibnamefont {Sleator}}, \bibinfo {author} {\bibfnamefont
  {J.~A.}\ \bibnamefont {Smolin}}, \ and\ \bibinfo {author} {\bibfnamefont
  {H.}~\bibnamefont {Weinfurter}},\ }\href {\doibase 10.1103/PhysRevA.52.3457}
  {\bibfield  {journal} {\bibinfo  {journal} {Phys. Rev. A}\ }\textbf {\bibinfo
  {volume} {52}},\ \bibinfo {pages} {3457} (\bibinfo {year}
  {1995})}\BibitemShut {NoStop}%
\bibitem [{\citenamefont {Jones}(2013)}]{Jones2013}%
  \BibitemOpen
  \bibfield  {author} {\bibinfo {author} {\bibfnamefont {C.}~\bibnamefont
  {Jones}},\ }\href@noop {} {\bibfield  {journal} {\bibinfo  {journal} {Phys.
  Rev. A}\ }\textbf {\bibinfo {volume} {87}},\ \bibinfo {pages} {022328}
  (\bibinfo {year} {2013})}\BibitemShut {NoStop}%
\bibitem [{\citenamefont {Eastin}(2012)}]{Eastin2012}%
  \BibitemOpen
  \bibfield  {author} {\bibinfo {author} {\bibfnamefont {B.}~\bibnamefont
  {Eastin}},\ }\href@noop {} {\enquote {\bibinfo {title} {Distilling one-qubit
  magic states into {T}offoli states},}\ } (\bibinfo {year} {2012}),\ \bibinfo
  {note} {\emph{Preprint} arXiv:1212.4872v1}\BibitemShut {NoStop}%
\bibitem [{\citenamefont {Raussendorf}\ \emph {et~al.}(2007)\citenamefont
  {Raussendorf}, \citenamefont {Harrington},\ and\ \citenamefont
  {Goyal}}]{Raussendorf2007}%
  \BibitemOpen
  \bibfield  {author} {\bibinfo {author} {\bibfnamefont {R.}~\bibnamefont
  {Raussendorf}}, \bibinfo {author} {\bibfnamefont {J.}~\bibnamefont
  {Harrington}}, \ and\ \bibinfo {author} {\bibfnamefont {K.}~\bibnamefont
  {Goyal}},\ }\href@noop {} {\bibfield  {journal} {\bibinfo  {journal} {New
  Journal of Physics}\ }\textbf {\bibinfo {volume} {9}},\ \bibinfo {pages}
  {199} (\bibinfo {year} {2007})}\BibitemShut {NoStop}%
\bibitem [{\citenamefont {Fowler}\ \emph {et~al.}(2009)\citenamefont {Fowler},
  \citenamefont {Stephens},\ and\ \citenamefont {Groszkowski}}]{Fowler2009}%
  \BibitemOpen
  \bibfield  {author} {\bibinfo {author} {\bibfnamefont {A.~G.}\ \bibnamefont
  {Fowler}}, \bibinfo {author} {\bibfnamefont {A.~M.}\ \bibnamefont
  {Stephens}}, \ and\ \bibinfo {author} {\bibfnamefont {P.}~\bibnamefont
  {Groszkowski}},\ }\href@noop {} {\bibfield  {journal} {\bibinfo  {journal}
  {Phys. Rev. A}\ }\textbf {\bibinfo {volume} {80}},\ \bibinfo {pages} {052312}
  (\bibinfo {year} {2009})}\BibitemShut {NoStop}%
\bibitem [{\citenamefont {Knill}(2004)}]{Knill2004}%
  \BibitemOpen
  \bibfield  {author} {\bibinfo {author} {\bibfnamefont {E.}~\bibnamefont
  {Knill}},\ }\href@noop {} {\enquote {\bibinfo {title} {Fault-tolerant
  postselected quantum computation: {S}chemes},}\ } (\bibinfo {year} {2004}),\
  \bibinfo {note} {\emph{Preprint} arXiv:quant-ph/0402171}\BibitemShut
  {NoStop}%
\bibitem [{\citenamefont {Bravyi}\ and\ \citenamefont
  {Kitaev}(2005)}]{Bravyi2005}%
  \BibitemOpen
  \bibfield  {author} {\bibinfo {author} {\bibfnamefont {S.}~\bibnamefont
  {Bravyi}}\ and\ \bibinfo {author} {\bibfnamefont {A.}~\bibnamefont
  {Kitaev}},\ }\href {\doibase 10.1103/PhysRevA.71.022316} {\bibfield
  {journal} {\bibinfo  {journal} {Phys. Rev. A}\ }\textbf {\bibinfo {volume}
  {71}},\ \bibinfo {pages} {022316} (\bibinfo {year} {2005})}\BibitemShut
  {NoStop}%
\bibitem [{\citenamefont {Meier}\ \emph {et~al.}(2012)\citenamefont {Meier},
  \citenamefont {Eastin},\ and\ \citenamefont {Knill}}]{Meier2012}%
  \BibitemOpen
  \bibfield  {author} {\bibinfo {author} {\bibfnamefont {A.~M.}\ \bibnamefont
  {Meier}}, \bibinfo {author} {\bibfnamefont {B.}~\bibnamefont {Eastin}}, \
  and\ \bibinfo {author} {\bibfnamefont {E.}~\bibnamefont {Knill}},\
  }\href@noop {} {\enquote {\bibinfo {title} {Magic-state distillation with the
  four-qubit code},}\ } (\bibinfo {year} {2012}),\ \bibinfo {note}
  {\emph{Preprint} arXiv:1204.4221v1}\BibitemShut {NoStop}%
\bibitem [{\citenamefont {Bravyi}\ and\ \citenamefont
  {Haah}(2012)}]{Bravyi2012}%
  \BibitemOpen
  \bibfield  {author} {\bibinfo {author} {\bibfnamefont {S.}~\bibnamefont
  {Bravyi}}\ and\ \bibinfo {author} {\bibfnamefont {J.}~\bibnamefont {Haah}},\
  }\href {\doibase 10.1103/PhysRevA.86.052329} {\bibfield  {journal} {\bibinfo
  {journal} {Phys. Rev. A}\ }\textbf {\bibinfo {volume} {86}},\ \bibinfo
  {pages} {052329} (\bibinfo {year} {2012})}\BibitemShut {NoStop}%
\bibitem [{\citenamefont {Jones}(2012)}]{Jones2012c}%
  \BibitemOpen
  \bibfield  {author} {\bibinfo {author} {\bibfnamefont {C.}~\bibnamefont
  {Jones}},\ }\href@noop {} {\enquote {\bibinfo {title} {Multilevel
  distillation of magic states for quantum computing},}\ } (\bibinfo {year}
  {2012}),\ \bibinfo {note} {\emph{Preprint} arXiv:1210.3388v1}\BibitemShut
  {NoStop}%
\bibitem [{\citenamefont {Gottesman}(1997)}]{Gottesman1997}%
  \BibitemOpen
  \bibfield  {author} {\bibinfo {author} {\bibfnamefont {D.}~\bibnamefont
  {Gottesman}},\ }\emph {\bibinfo {title} {Stabilizer Codes and Quantum Error
  Correction}},\ \href@noop {} {Ph.D. thesis},\ \bibinfo  {school} {California
  Institute of Technology}, \bibinfo {address} {Pasadena, CA} (\bibinfo {year}
  {1997})\BibitemShut {NoStop}%
\bibitem [{\citenamefont {Fowler}\ and\ \citenamefont
  {Devitt}(2012{\natexlab{a}})}]{Fowler2012d}%
  \BibitemOpen
  \bibfield  {author} {\bibinfo {author} {\bibfnamefont {A.~G.}\ \bibnamefont
  {Fowler}}\ and\ \bibinfo {author} {\bibfnamefont {S.~J.}\ \bibnamefont
  {Devitt}},\ }\href@noop {} {\enquote {\bibinfo {title} {A bridge to lower
  overhead quantum computation},}\ } (\bibinfo {year} {2012}{\natexlab{a}}),\
  \bibinfo {note} {\emph{Preprint} arXiv:1209.0510v3}\BibitemShut {NoStop}%
\bibitem [{\citenamefont {Fowler}\ and\ \citenamefont
  {Devitt}(2012{\natexlab{b}})}]{Fowler2012}%
  \BibitemOpen
  \bibfield  {author} {\bibinfo {author} {\bibfnamefont {A.~G.}\ \bibnamefont
  {Fowler}}\ and\ \bibinfo {author} {\bibfnamefont {S.~J.}\ \bibnamefont
  {Devitt}},\ }\href@noop {} {\enquote {\bibinfo {title} {A bridge to lower
  overhead quantum computation},}\ } (\bibinfo {year} {2012}{\natexlab{b}}),\
  \bibinfo {note} {\emph{Preprint} arXiv:1209.0510v3}\BibitemShut {NoStop}%
\bibitem [{\citenamefont {Fowler}\ \emph {et~al.}(2013)\citenamefont {Fowler},
  \citenamefont {Devitt},\ and\ \citenamefont {Jones}}]{Fowler2013}%
  \BibitemOpen
  \bibfield  {author} {\bibinfo {author} {\bibfnamefont {A.~G.}\ \bibnamefont
  {Fowler}}, \bibinfo {author} {\bibfnamefont {S.~J.}\ \bibnamefont {Devitt}},
  \ and\ \bibinfo {author} {\bibfnamefont {C.}~\bibnamefont {Jones}},\
  }\href@noop {} {\enquote {\bibinfo {title} {Surface code implementation of
  block code state distillation},}\ } (\bibinfo {year} {2013}),\ \bibinfo
  {note} {\emph{Preprint} arXiv:1301.7107v1}\BibitemShut {NoStop}%
\bibitem [{\citenamefont {DiVincenzo}\ and\ \citenamefont
  {Aliferis}(2007)}]{DiVincenzo2007}%
  \BibitemOpen
  \bibfield  {author} {\bibinfo {author} {\bibfnamefont {D.~P.}\ \bibnamefont
  {DiVincenzo}}\ and\ \bibinfo {author} {\bibfnamefont {P.}~\bibnamefont
  {Aliferis}},\ }\href@noop {} {\bibfield  {journal} {\bibinfo  {journal}
  {Phys. Rev. Lett.}\ }\textbf {\bibinfo {volume} {98}},\ \bibinfo {pages}
  {020501} (\bibinfo {year} {2007})}\BibitemShut {NoStop}%
\bibitem [{\citenamefont {Aliferis}(2007)}]{Aliferis_thesis}%
  \BibitemOpen
  \bibfield  {author} {\bibinfo {author} {\bibfnamefont {P.}~\bibnamefont
  {Aliferis}},\ }\emph {\bibinfo {title} {{Level Reduction and the Quantum
  Threshold Theorem}}},\ \href@noop {} {Ph.D. thesis},\ \bibinfo  {school}
  {California Institute of Technology} (\bibinfo {year} {2007})\BibitemShut
  {NoStop}%
\bibitem [{\citenamefont {Gottesman}\ and\ \citenamefont
  {Chuang}(1999)}]{Gottesman1999}%
  \BibitemOpen
  \bibfield  {author} {\bibinfo {author} {\bibfnamefont {D.}~\bibnamefont
  {Gottesman}}\ and\ \bibinfo {author} {\bibfnamefont {I.~L.}\ \bibnamefont
  {Chuang}},\ }\href@noop {} {\bibfield  {journal} {\bibinfo  {journal}
  {Nature}\ }\textbf {\bibinfo {volume} {402}},\ \bibinfo {pages} {390}
  (\bibinfo {year} {1999})}\BibitemShut {NoStop}%
\bibitem [{\citenamefont {Shor}(1997)}]{Shor1999}%
  \BibitemOpen
  \bibfield  {author} {\bibinfo {author} {\bibfnamefont {P.~W.}\ \bibnamefont
  {Shor}},\ }\href@noop {} {\bibfield  {journal} {\bibinfo  {journal} {SIAM J.
  Comput.}\ }\textbf {\bibinfo {volume} {26}},\ \bibinfo {pages} {1484}
  (\bibinfo {year} {1997})}\BibitemShut {NoStop}%
\bibitem [{\citenamefont {Lloyd}(1996)}]{Lloyd1996}%
  \BibitemOpen
  \bibfield  {author} {\bibinfo {author} {\bibfnamefont {S.}~\bibnamefont
  {Lloyd}},\ }\href {\doibase 10.1126/science.273.5278.1073} {\bibfield
  {journal} {\bibinfo  {journal} {Science}\ }\textbf {\bibinfo {volume}
  {273}},\ \bibinfo {pages} {1073} (\bibinfo {year} {1996})}\BibitemShut
  {NoStop}%
\bibitem [{\citenamefont {Jones}\ \emph
  {et~al.}(2012{\natexlab{b}})\citenamefont {Jones}, \citenamefont {Whitfield},
  \citenamefont {McMahon}, \citenamefont {Yung}, \citenamefont {Meter},
  \citenamefont {Aspuru-Guzik},\ and\ \citenamefont {Yamamoto}}]{Jones2012b}%
  \BibitemOpen
  \bibfield  {author} {\bibinfo {author} {\bibfnamefont {N.~C.}\ \bibnamefont
  {Jones}}, \bibinfo {author} {\bibfnamefont {J.~D.}\ \bibnamefont
  {Whitfield}}, \bibinfo {author} {\bibfnamefont {P.~L.}\ \bibnamefont
  {McMahon}}, \bibinfo {author} {\bibfnamefont {M.-H.}\ \bibnamefont {Yung}},
  \bibinfo {author} {\bibfnamefont {R.~V.}\ \bibnamefont {Meter}}, \bibinfo
  {author} {\bibfnamefont {A.}~\bibnamefont {Aspuru-Guzik}}, \ and\ \bibinfo
  {author} {\bibfnamefont {Y.}~\bibnamefont {Yamamoto}},\ }\href@noop {}
  {\bibfield  {journal} {\bibinfo  {journal} {New Journal of Physics}\ }\textbf
  {\bibinfo {volume} {14}},\ \bibinfo {pages} {115023} (\bibinfo {year}
  {2012}{\natexlab{b}})}\BibitemShut {NoStop}%
\bibitem [{\citenamefont {Aliferis}\ and\ \citenamefont
  {Cross}(2007)}]{Aliferis2007}%
  \BibitemOpen
  \bibfield  {author} {\bibinfo {author} {\bibfnamefont {P.}~\bibnamefont
  {Aliferis}}\ and\ \bibinfo {author} {\bibfnamefont {A.~W.}\ \bibnamefont
  {Cross}},\ }\href {\doibase 10.1103/PhysRevLett.98.220502} {\bibfield
  {journal} {\bibinfo  {journal} {Phys. Rev. Lett.}\ }\textbf {\bibinfo
  {volume} {98}},\ \bibinfo {pages} {220502} (\bibinfo {year}
  {2007})}\BibitemShut {NoStop}%
\end{thebibliography}

%

\end{document}